\documentclass[preprint,nofootinbib,aps,showpacs,preprintnumbers,amsmath,amssymb]{revtex4}
\usepackage{graphicx}% Include figure files
\usepackage{dcolumn}% Align table columns on decimal point
\usepackage{bm}% bold math
\usepackage{natbib}

\newcommand{\beq}{\begin{equation}}
\newcommand{\eq}{\end{equation}}
\newcommand{\bea}{\begin{eqnarray}\displaystyle}
\newcommand{\ea}{\end{eqnarray}}

\newcommand{\s}{\sigma}
\newcommand{\p}{\partial}

\newcommand{\e}{\epsilon}
\newcommand{\Xp}{X_+}
\newcommand{\Xm}{X_-}
\begin{document}
\preprint{Brown-HET-1453}

\title{Tachyon Backgrounds in 2D String Theory}

\author{Sera Cremonini}

\email{sera@het.brown.edu}
\affiliation{Department of Physics, Brown University \\ Providence,
RI 02912, USA}
\date{\today}

\begin{abstract}
We consider the construction of tachyonic backgrounds in
two-dimensional string theory, focusing on the Sine-Liouville
background. This can be studied in two different ways, one within
the context of collective field theory and the other via
the formalism of Toda integrable systems. The two approaches are
seemingly different. The latter involves a deformation of the
original inverted oscillator potential while the former does not.
We perform a comparison by explicitly constructing
the Fermi surface in each case, and demonstrate that the two
apparently different approaches are in fact equivalent.
\end{abstract}

\maketitle

\section{Introduction}

Recent developments in string theory have taught us the importance
of holography in trying to understand gravity and its relationship
to conformal field theories (CFTs).
A prime example of the duality between gravitational physics and a
lower-dimensional theory not containing gravity is given by the AdS/CFT correspondence
\cite{Maldacena:1997re}.
However, the earliest example of a theory admitting such a holographic description is
two-dimensional (2D) non-critical string theory
(see \cite{Klebanov:1991qa,Ginsparg:1993is,Jevicki:1993qn,Polchinski:1994mb,Martinec:2004td} for reviews).
In this case the duality is between one-dimensional $c=1$ matrix
quantum mechanics and two-dimensional string theory, realized as a collective field
theory. This has also been understood as an example of open-closed
string duality \cite{McGreevy:2003kb,Klebanov:2003km,Sen:2003iv}.

Although 2D string theory is just a toy model, it has
proven to be very useful because of its solvability. It is, in
fact, an example of a sigma-model with integrable dynamics.
It also allows us to address other relevant string theory issues, such as time dependence and
the search for new backgrounds, within the context of a model that
is more tractable.

In 2D string theory the only dynamical degree of
freedom is the massless scalar tachyon, whose dynamics (scattering
amplitudes) has been shown to be well described by the collective field theory of a
single scalar \cite{Das:1990ka}. Furthermore, the effective (collective) field
theory of the tachyon is exactly solvable \cite{Polchinski:1991uq}.

Most of the 2D string theory investigations have involved
the so-called Liouville background, described in the CFT by a
$c=1$ matter field coupled to the Liouville field.
Additional backgrounds can be obtained by replacing the Liouville
term by a general tachyon (vertex) perturbation. Such perturbed
backgrounds have been studied within the context of collective field theory
\cite{Lee:1993fa} as well as by Alexandrov, Kazakov and Kostov (AKK) via the formalism of Toda
integrable systems \cite{Alexandrov:2002fh,Kostov:2002tk}.
Also, a lot of effort has been recently devoted to the study of
cosmological issues in certain time-dependent backgrounds
\cite{Karczmarek:2003pv,Karczmarek:2004ph,Das:2004hw,Karczmarek:2004yc,Ernebjerg:2004ut,Mukhopadhyay:2004ff,Das:2004aq,Das:2005jp}.

In this work we will concentrate on the Sine-Liouville background, as an example of the tachyon
perturbations mentioned above, and we will compare the two seemingly different ways in which it has been
addressed. In the collective field approach of \cite{Lee:1993fa}, the Hamiltonian was that of the standard
inverted harmonic oscillator, while in the work of AKK the Sine-Lioville potential was simulated by introducing certain {\it deformed} Hamiltonians.
At the moment it is not clear what the correspondence between these two
different approaches is. It is the purpose of this paper to
clarify this issue.
In particular, we will explicitly construct the Fermi
surface for each distinct approach, and show that the two methods are equivalent.
We should mention that some of the simpler backgrounds which
were recently studied in
\cite{Karczmarek:2003pv,Karczmarek:2004ph,Das:2004hw,Karczmarek:2004yc,Ernebjerg:2004ut,Mukhopadhyay:2004ff,Das:2004aq,Das:2005jp}
were explicitly shown to arise as solutions of the standard
inverted oscillator collective field theory.

The paper is organized as follows.
In Section II we describe certain 2D string theory backgrounds,
and we introduce the Sine-Liouville model as a perturbation of Liouville theory.
Section III contains the basics of collective field theory, as
well as the collective approach to building non-trivial backgrounds.
This construction is then applied to the case of the Sine-Liouville
background, and it is used to obtain the explicit shape of the
corresponding Fermi surface.
Section IV outlines the method of \cite{Alexandrov:2002fh,Kostov:2002tk} for constructing new
backgrounds. This serves as a starting point for extracting
the Sine-Liouville Fermi surface, but in a manner entirely different from
that of the collective field theory.
Section V consists of a comparison of the two approaches, and shows agreement of the Fermi surfaces.

%%%%%%%%%%%%%%%%%%%%%%%%%%%%%%%%%%%%%%%%%%%%%%%%%%%%%%%%%%%%%%%%%%%%%%%%

\section{2D string theory backgrounds \label{2DSect}}

%%%%%%%%%%%%%%%%%%%%%%%%%%%%%%%%%%%%%%%%%%%%%%%%%%%%%%%%%%%%%%%%%%%%%%%%

One way to discuss the dynamics of strings is through the
$\beta$-function approach, which provides the effective field
theory description of the low-energy fields.
We would like to start by reviewing the connection between the
nonlinear $\s$-model and the corresponding effective theory.

The 2D nonlinear $\s$-model is given by
\beq
S_\s = \frac{1}{4\pi} \int d^2\sigma \sqrt{g} \, [g^{a b} \p_a X^\mu \p_b X^\nu G_{\mu \nu}(X) + R^{(2)}D(X) + T(X) + ...]
\eq
where $X \equiv X^\mu = (X^0,\varphi)$ parametrize the two-dimensional
target space, $T(X)$ is the massless tachyon, $D(X)$ is the dilaton and $G_{\mu \nu}(X)$ the
graviton.
The vanishing of the $\beta$-functions guarantees conformal
invariance, and yields the spacetime equations of motion.
As is well known, the $\beta$-function equations can be shown to follow
from the action
\beq
S_{eff} = \frac{1}{2\pi \kappa^2} \int d^2 X \sqrt{G} e^{-2D(X)} \left[R+ 4
(\nabla
D)^2 - (\nabla T)^2 +2 T^2 -V(T)\right],
\eq
where $\kappa$ is the string coupling and $V(T)$ is the tachyon potential, which
includes tachyon interactions and which we leave
unspecified.

One solution ensuring the vanishing of the $\beta$-functions is the (Euclidean) {\it linear dilaton
vacuum}, given by
\bea
G_{\mu \nu}&=&\delta_{\mu \nu}, \nonumber \\
D(X)&=&\sqrt{2} \varphi, \nonumber \\
T(X)&=&0.
\ea
In order to find more general solutions, let us consider the
{\it linearized} equation of motion for the tachyon,
\beq
\label{Teom}
\beta^T=-2\nabla^2 T + 4 \nabla D \nabla T -4 T=0.
\eq
With the above choice for the dilaton, $D(X)=\sqrt{2} \varphi$, eq.(\ref{Teom}) becomes
\beq
\label{Teom2}
(\p_{X^0}^2+\p_{\varphi}^2-2\sqrt{2}\p_{\varphi}+2)T=0.
\eq
The linearized {\it static} tachyon equation $(\p_{\varphi}^2-2\sqrt{2}\p_{\varphi}+2)T=0$ has two linearly
independent solutions, $e^{\sqrt{2}\varphi}$ and $\varphi e^{\sqrt{2}\varphi}$.
One can show that by taking
\bea
G_{\mu \nu}&=&\delta_{\mu \nu}, \nonumber \\
D(X)&=&\sqrt{2} \varphi, \nonumber \\
T(X)&=& \frac{\mu}{4} e^{\sqrt{2} \varphi}
\ea
we obtain another consistent string theory solution. This choice specifies the
so-called Liouville background,
\beq
S_0 = \frac{1}{8\pi}\int d^2\s \left[ (\p X^0)^2 +(\p \varphi)^2 + 2\sqrt{2} \varphi R^{(2)} + \mu  e^{\sqrt{2} \varphi}
\right],
\eq
a $c=1$ conformal field theory coupled to the Liouville field
$\varphi$. The central charge $c\equiv c_X=1$ refers to the matter
coordinate $X^0$, while the Liouville field with $Q=2\sqrt{2}$ carries a central charge
$c_\varphi=1+3Q^2=25$, giving us the expected total charge
$c_{X^0}+c_\varphi=26$.
So far we have only considered static solutions of the tachyon
linearized equation. If we allow for time dependence, we can see
that tachyons of the form $T=e^{(\sqrt{2}\pm  p)\varphi}\, e^{\pm i p
X^0}$ are also solutions of (\ref{Teom}).
In particular, the specific linear combination $T=e^{(\sqrt{2}- |p|)\varphi}\,
\cos(pX^0)$ is what is usually referred to as the Sine-Liouville
background.

Liouville theory can also be obtained in the following way.
Start from the Euclidean string action
\beq
S_0 \sim \int d^2 \sigma \sqrt{detg} \bigl(g^{ab} \partial_a X^0 \partial_b X^0 +\mu
\bigr),
\eq
with $g^{ab}$ the world-sheet metric, $\mu$ the ``cosmological constant'' and
$X(\sigma)$ the embedding of the string into Euclidean time.
Choosing the conformal gauge $g^{ab}=e^\varphi \hat{g}^{ab}$,  the dilaton $\varphi$ becomes dynamical
(due to the conformal anomaly),
and the world-sheet CFT action takes the Liouville form
\beq
S_0 \sim \int d^2 \sigma \bigl[(\partial X^0)^2 + (\partial \varphi)^2 + Q R^{(2)} \varphi + \mu e^{\gamma\varphi}
\bigr].
\eq
The background charge $Q$ is determined by the usual
central charge requirement $c_\varphi=1+3Q^2=25$, while the Liouville exponent $\gamma$
is found by imposing that the cosmological term perturbation is
marginal, $-\frac{1}{2}\gamma(\gamma-Q)=1.$
From this perspective, the Liouville interaction term is thought of as the
insertion of a vertex operator, $\delta L = \mu
V_\gamma$, with $ V_\gamma=e^{\gamma \varphi}$.

More general perturbations can be considered.
In fact, if the Euclidean time is compactified with radius $R$, the
spectrum of admissible momenta becomes discrete,
$p_n=n/R$, $n \, \epsilon \, \mathbb{Z}$. In addition, there is also a discrete spectrum
of operators describing winding modes on the world-sheet, with charges $q_m=m R$, $m \, \epsilon \, \mathbb{Z}$.
The form of {\it vertex} operators $V_p$ describing momentum modes, and {\it vortex} operators $\tilde{V}_q$
describing winding modes is
\bea
V_p & \sim &  \int d^2 \sigma e^{ipX_0} e^{(\sqrt{2}-|p|)\varphi},\\
\tilde{V}_q & \sim & \int d^2 \sigma e^{iq\tilde{X}_0}
e^{(\sqrt{2}-|q|)\varphi},
\ea
where $\tilde{X}^0$ is the dual coordinate to $X^0$.
One can obtain new backgrounds of the compactified 2D string
theory by perturbing the Liouville action with both vertex and
vortex operators:
\beq
S=S_0+\sum_{n \neq 0}(t_n V_n + \tilde{t}_n \tilde{V}_n).
\eq
In particular, considering only momentum mode perturbations and
choosing $n=\pm 1$, we reproduce the Sine-Liouville background (with $p_1=\frac{1}{R}$) :
\beq
S \sim \int d^2\sigma \bigl[(\partial X^0)^2 + (\partial \varphi)^2 +Q R^{(2)} \varphi +
\mu e^{\gamma\varphi} +\lambda
e^{(\sqrt{2}-\frac{1}{R})\varphi}\cos(\frac{X^0}{R})\bigr],
\eq
where the couplings have been taken to be
$t_1=t_{-1}=\frac{\lambda}{2}$.

%%%%%%%%%%%%%%%%%%%%%%%%%%%%%%%%%%%%%%%%%%%%%%%%%%%%%%%%%%%%%%%%%%%%%%%%%
\section{The collective field theory approach}
%%%%%%%%%%%%%%%%%%%%%%%%%%%%%%%%%%%%%%%%%%%%%%%%%%%%%%%%%%%%%%%%%%%%%%%%%

A correspondence between one-dimensional matrix quantum mechanics
and two-dimensional string theory can be established by using the formalism of collective field theory
(see \cite{Jevicki:1993qn} for a review).
Consider the theory of a hermitian $N\times N$ matrix $M(t)$ in one time
dimension, with Lagrangian
\beq
L=\frac{1}{2}\, \text{Tr} \left(\dot{M}^2-u(M)\right).
\eq
The non-critical $c=1$ string theory corresponds to choosing
an inverted harmonic oscillator potential, $u(M)=-\frac{M^2}{2}$.
There is an associated $U(N)$ conserved charge
$J=i[M,\dot{M}].$
In the singlet sector $J | \; \rangle = 0$ the model becomes a
gauge theory, and the matrix $M$ can be diagonalized, $M(t) \rightarrow
diag(\lambda_i(t)).$
It can be shown that the $U(N)$ invariant sector of the matrix model is described
by a non-relativistic quantum mechanics of free fermions - the eigenvalues of the
matrix.
Collective variables for the model are introduced in the following
way:
\beq
\phi_k (t) = \text{Tr} \left(e^{ikM(t)} \right) = \sum_{i=1}^N
e^{ik\lambda_i(t)}\, .
\eq
Notice that so far we have been dealing with a one-dimensional
(matrix) theory, depending only on time.
A two-dimensional model arises naturally when one introduces an
additional spatial dimension $x$, which is related to the space of
eigenvalues $\lambda_i$ of the matrix $M(t)$:
\beq
\phi(x,t)=\frac{1}{2\pi} \int dk \; e^{-i k x} \, \phi_k(t) =
\sum_{i=1}^N\delta(x-\lambda_i(t)).
\eq
Thus, the collective field has the physical interpretation of a density field of
fermions.
Finally, introduction of a conjugate field $\Pi_\phi$ with Poisson
brackets $\{\phi(x) ,\Pi_\phi(y)\}=\delta(x-y)$
yields a canonical phase space.

One important difference between the matrix model and the
string theory side which will be evident soon is that, while the collective equations of
motion are highly nonlinear, the matrix equations (for the oscillator potential) are linear.
This feature of integrability, or exact solvability, is one of the main advantages of
applying the matrix model formalism to the study of
2D string theories. It is precisely through the nonlinear transformation
$\phi(x,t)=\text{Tr} \delta(x-M(t))$ that the
matrix model provides an exact solution to the corresponding nonlinear field theory.

The dynamics of the collective field theory can be induced
from that of the matrix model variables $M(t)$ and $P(t)=\dot{M}(t)$.
At the classical level the collective action is given by
\beq
S= \int dx \, dt \left[ \frac{1}{2}\frac{(\p^{-1}_x
\dot{\phi})^2}{\phi}-\frac{\pi^2}{6}\phi^3+\frac{1}{2}(x^2+\mu)\phi
\right].
\eq
The cosmological constant $\mu$ is introduced as a
``chemical potential'' coupled to the size of the matrix. It acts
as a Lagrange multiplier enforcing the normalization condition
$\int dx \, \phi(x,t) = N$.
We can eliminate the explicit $\mu$ dependence, which we do by properly
rescaling $x$ and $\phi$,
$ x \rightarrow  g^{-1/2}x, \; \; \phi \rightarrow g^{-1/2}\phi$,
and setting $g \mu =1$.

The collective field $\phi(x,t)$ and its
conjugate momentum $\Pi_\phi$ can be written in terms of left- and
right-moving fields $p_\pm(x,t)$,
\beq p_\pm (x,t)= g^2 \partial_x
\Pi_\phi(x,t) \pm \pi \phi(x,t),
\eq
in terms of which the action
becomes
\beq
S=-\frac{1}{2\pi g^2}\int dx dt \left[\frac{1}{2}
\Bigl(p_+ \p_x^{-1} \dot{p}_+-p_- \p_x^{-1} \dot{p}_- \Bigr)
+\frac{1}{6}
\Bigl(p_+^3-p_-^3\Bigr)-\frac{1}{2}(x^2+1)\Bigl(p_+-p_-\Bigr)
\right]. \nonumber
\eq
The equation of motion for the fields $p_\pm$ is
\beq
\label{eom}
\p_t p_\pm = x-p_\pm \p_x p_\pm
\eq
whose static solution is given by $p_\pm = \pm \sqrt{x^2+1}.$
We expand the fields around the static background,
\beq
p_\pm =\pm
\sqrt{x^2+1}+\eta_\pm,
\eq
where the fluctuations $\eta_\pm$ are in general
time-dependent.
It will be convenient to introduce a new
coordinate $\tau$ such that $x=\sinh(\tau)$, and define $\alpha_\pm
= \eta_\pm \cosh(\tau)$. The action $S=S^+ + S^-$ then becomes
\beq
S^{\pm}=\mp \frac{1}{4\pi} \int dt d\tau \left[-\alpha_\pm
\p_\tau^{-1}\dot{\alpha}_\pm +\alpha_\pm^2 + \frac{g}{3} \frac{\alpha_\pm
^3}{\cosh(\tau)^2} \right].
\eq
Since $\alpha_\pm$ are decoupled,
it is enough to study only one of them; we will consider $\alpha \equiv
\alpha^+$. Its equation of motion is
\beq
(\p_t+\p_\tau)\alpha =
-\frac{g}{2}\p_\tau \left(\frac{\alpha^2}{ \cosh(\tau)^2}\right),
\eq
which can be solved perturbatively in $g$.
Let $\alpha_0$ be
the solution of the free equation, $(\p_t+\p_\tau)\alpha_0 =0$;
one can then solve for $\alpha$ in terms of the linearized
field $\alpha_0$,
\beq
\alpha=\alpha_0+g \, \alpha_1 + g^2\,
\alpha_2 + ...
\eq
Note that $\alpha$ and $\alpha_0$ are connected through the relation
\beq
\alpha_0(t,\tau)=\alpha(t,\tau)-\int \, dt^\prime \, d\tau^\prime \,
\Delta_F(t-t^\prime,\tau-\tau^\prime)\, (\p_{t^\prime}+\p_{\tau^\prime}) \,
\alpha(t^\prime,\tau^\prime),
\eq
where $\Delta_F$ is the Feynman propagator.
It will also turn out to be useful to introduce in and out fields
$\alpha_{in,\, out}$
\beq
\alpha_{in,\,out}(t,\tau)=\alpha(t,\tau)-\int \, dt^\prime \, d\tau^\prime \,
\Delta_{R,\,A}(t-t^\prime,\tau-\tau^\prime)\, (\p_{t^\prime}+\p_{\tau^\prime}) \,
\alpha(t^\prime,\tau^\prime),
\eq
with $\Delta_{R,\,A}$ retarded and advanced propagators, and to
note that
\bea
\label{alpha1}
\lim_{t\rightarrow -\infty} \alpha(t,\tau) &=& \lim_{t\rightarrow -\infty}
\alpha_{in}(t,\tau) \nonumber \\
\lim_{t\rightarrow \infty} \alpha(t,\tau) &=& \lim_{t\rightarrow \infty}
\alpha_{out}(t,\tau).
\ea
Finally, using properties of propagators, one can show that the
Fourier transforms of these fields are related by
\beq
\label{alpha2}
\tilde{\alpha}_0(w,k)=\theta(k) \tilde{\alpha}_{out}(w,k)+\theta(-k)
\tilde{\alpha}_{in}(w,k).
\eq
The relations above are important because they uniquely
determine $\alpha$ in terms of $\alpha_{0}$; furthermore, any one of $\alpha_0$, $\alpha_{in}$ or $\alpha_{out}$
contains all the information about the boundary values of the system.

It is precisely the linearized collective field
$\alpha_0$ that is identified with the world-sheet tachyon $T(X_0,\varphi)$.
A quick way to see this is the following. Note that $(\p_t+\p_\tau)\alpha_0 =0$ can be
rewritten as $(-\p^2_t+\p^2_\tau)\alpha_0 =0$.
Also, recall that the tachyon obeyed the linearized
equation
\beq
(\p^2_{X_0}+\p^2_\varphi-2\sqrt{2}\p_\varphi+2)T=0.
\eq
After defining a new field $S$ given by $T=e^{\sqrt{2}\varphi}S$, the
above equation of motion becomes
\beq
(-\p_{t}^2+\p_{\varphi}^2)S=0, \;\;\; \text{after} \;\;\;  X_0=i t,
\eq
showing that the rescaled tachyon $S$ obeys the same equation as
$\alpha_0$.
Thus, the relevant collective field theory quantity which we need
to identify when trying to make a connection with a given tachyon
background is going to be $\alpha_0$.

An {\it exact} solution of the nonlinear
classical equation (\ref{eom}) was obtained in parametric form in
\cite{Polchinski:1991uq}. Using $\s$ to parametrize $x$ and $p$, we have
\bea
x&=&a(\sigma)\sinh(t-\sigma), \nonumber \\
p&=&a(\sigma)\cosh(t-\sigma), \nonumber
\ea
obeying
\beq \label{FS}
p^2-x^2=a^2(\s)=1-g\, \e(\s),
\eq
where the new field $\e(\s)$ denotes fluctuations about the static background.
It is precisely this equation which specifies the profile of the Fermi sea, {\it
i.e.} the Fermi surface.
We can see from (\ref{FS}) that it is $\e(\s)$ which encodes the essential
information about a given background;
thus, specifying $\e(\s)$ corresponds to specifying the background completely.
Since, as we have just seen, the world-sheet tachyon T can be related to
$\alpha_0$, in order to find the Fermi surface
corresponding to a given background, we need
to express $\e(\s)$ in terms of $\alpha_0$. This was done in
\cite{Lee:1993fa}. We will briefly review the
arguments of that work.

We start by considering a point on the Fermi surface with coordinate $x$. We evolve
it back in time to $t\rightarrow - \infty$, and turn off the
interaction, by setting $g=0$. We then follow it back to
present time, and label it by $x_{in}$. Since
$x=\sinh(\tau)=\sqrt{1-g\e}\sinh(t-\s)$, we also have
$x_{in}=\sinh(\tau_{in})=\sinh(t-\tilde{\s})$.
Combining these two expressions gives the equation
\beq
\tau (t=-\infty,\s)=\tau_{in}(t=-\infty,\tilde{\s}),
\eq
which implies
\beq
\tilde{\s}=\s + \ln(1-g\e)/2.
\eq
This gives
\beq
\label{TauIn}
\tau_{in}(t,\tilde{\s})=t-\tilde{\s}=t-\s-\frac{1}{2}\ln(1-g\e(\s)).
\eq
In the $t \rightarrow -\infty$ limit, it can be shown \cite{Lee:1993fa} that $\alpha
= -\frac{\e(\s)}{2}+ {\cal O}(e^{2(t-\s)})$.
This allows us to write, using (\ref{alpha1}),
\bea
\e(\s) &=& - 2 \lim_{t\rightarrow -\infty} \alpha(t,\tau) = - 2 \lim_{t\rightarrow -\infty}
\alpha_{in}(t-\tau_{in}) \nonumber \\
&=& - 2 \alpha_{in}(\s+\frac{1}{2}\ln(1-g\e(\s))) =
\e_{in}(\s+\frac{1}{2}\ln(1-g\e(\s))),
\ea
where we have used (\ref{TauIn}) and defined $\e_{in}\equiv - 2 \alpha_{in}$.
Similarly one can obtain an expression for $\e(\s)$ in terms of $\e_{out} \equiv -2
\alpha_{out}$,
\beq
\e(\s) = \e_{out}(\s-\frac{1}{2}\ln(1-g\e(\s))).
\eq
These two expressions can be combined to directly relate $\e_{in}$ and $\e_{out}$:
\beq
\e_{in}(\s) = \e_{out}(\s-\ln(1-g\e_{in}(\s))).
\eq
The solution of the equation above is
\beq
\label{sol}
\tilde{\e}_{in,out}(k)=-\frac{1}{2\pi g} \frac{1}{1\mp i k/2}
\int_{-\infty}^{\infty} d\s e^{-ik\s}[(1-g\e(\s))^{1\mp i k/2}-1]
\eq
with $\tilde{\e}_{in,\, out}(k)$ denoting Fourier transforms.
This formula summarizes the strategy for
constructing non-trivial backgrounds.
Classical solutions can be characterized by their asymptotic
behavior, which corresponds to specification of a linearized
solution. In the above integral formula, this is equivalent to
specifying $\tilde{\e}_{in,\, out}(k)$. The exact solution can
then be found by solving for $\e(\s)$.
Next, we will focus on the special case of the Sine-Liouville background.

\subsection{The Sine-Liouville background}

So far the analysis has been general. To apply it to
the Sine-Liouville background, we need to specify
$\tilde{\e}_{in,out}(k)$.
Define the Fourier transform $\tilde{T}(k)$ of the tachyon field
in the following way,
\beq
T(X)=\int d k \, \tilde{T}(k) e^{i k X + \beta_k \varphi},
\eq
where $\beta_p \equiv \sqrt{2}(1-\frac{|p|}{2})$. \footnote{Notice that
the fields $X^0$ and $\varphi$ are related to those of Sect. \ref{2DSect} by simple rescalings.}
In \cite{Lee:1993fa} it is shown that $\tilde{T}(k)$ should be
identified with the Fourier transform of the collective field,
$\tilde{\alpha}_0(k)$.

Plugging in the expression for the Sine-Liouville tachyon, $T(X_0,\varphi)= \lambda e^{\beta_p \varphi} \cos(pX)$,
one finds
\beq
\tilde{\alpha}_0(k)= \frac{\lambda}{2}(\delta(k+p) +\delta(k-p)).
\eq
However, in order to allow for more general linear
combinations of the linearized solutions, we take
\beq
\tilde{\e}_0(k)\equiv \tilde{\alpha}_0(k)= \alpha \delta(k+p) +\beta
\delta(k-p).
\eq
Finally, using (\ref{alpha2}),
one gets
\bea
\tilde{\e}_{in}(k) &=& f(k)\theta(k) + \alpha \delta(k+p) \\
\tilde{\e}_{out}(k) &=& f(k)\theta(-k) + \beta \delta(k-p).
\ea
The unknown functions $f(k)$ and $\e(\sigma)$ can be solved by
plugging $\tilde{\e}_{in, \, out}(k)$ into (\ref{sol}).
This was done by \cite{Lee:1993fa} to several orders in perturbation
theory, but here we will stop at cubic order:
\bea
\e_0 &=& \alpha e^{-k\s}+ \beta e^{k\s}, \\
\e_1 &=& \frac{k}{2}(\alpha^2 e^{-2k\s} + \beta^2 e^{2k\s}), \\
\e_2 &=& (\frac{3k^2}{8}+\frac{k}{4})(\alpha^3 e^{-3k\s}+ \beta^3 e^{3k\s})+(\frac{k^2}{8}+\frac{k}{4})(\alpha^2 \beta e^{-k\s}+ \beta^2 \alpha
e^{k\s}).
\ea
By making use of this result, we will now explicitly construct the
Fermi surface.

%%%%%%%%%%%%%%%%%%
\subsection{Constructing the Fermi surface}
%%%%%%%%%%%%%%%%%%

%%%% NB: In this section the constants c_1 and c_2 were originally labeled c_2 and c_3 %%%%%%%%%%

Plugging $\e(\sigma)$ above into the generic Fermi surface (\ref{FS}) we find
\bea
\label{a}
a &\sim& 1-\frac{g}{2}\left[\alpha e^{-k\s} +\beta e^{k\s}\right]
-\frac{g^2}{8} \left[(1+2k)\bigl(\alpha^2 e^{-2k\s} +\beta^2 e^{2k\s}\bigr)+ 2\alpha\beta \right]+ \nonumber \\
&+& g^3 \left[ (-\frac{c_2}{2}-\frac{k}{8}-\frac{3}{16})\bigl(e^{-k\s}\alpha^2\beta+
e^{k\s}\alpha\beta^2\bigr)+(-\frac{c_1}{2}-\frac{k}{8}-\frac{1}{16})\bigl(\alpha^3 e^{-3k\s}+\beta^3
e^{3k\s}\bigr)\right] \nonumber
\\ &+& {\cal O}(g^4),
\ea
with
$c_1=\frac{3k^2}{8}+\frac{k}{4}$, $c_2=\frac{k^2}{8}+\frac{k}{4}$.
In order to express (\ref{FS}) entirely in terms of $p$ and $x$,
we need to find $\s=\s(p,x)$.
It turns out to be more convenient to work in light-cone
coordinates, which are typically taken to be
\bea
x_+&=&p+x=a(\sigma)e^{\, t-\s}, \nonumber \\
x_-&=&p-x=a(\sigma)e^{-t+\s}. \nonumber
\ea
However, in view of the comparison with the work of \cite{Alexandrov:2002fh,Kostov:2002tk},
we choose to incorporate the explicit time dependence
into the definition of new light-cone coordinates $X_\pm$, given by
\bea
\label{Xp}
X_+&=&e^{-t}x_+=a(\sigma)e^{-\s},  \\
\label{Xm}
X_-&=&e^{\, t}x_-=a(\sigma)e^{\, \s}.
\ea

Notice from (\ref{Xp}), (\ref{Xm}) that one can solve for $\sigma$ as a function of
either $X_+$ or $X_-$; this is a simple restatement of the fact that $p(x)$ is a two-valued
function (with two branches usually denoted by $p_\pm(x)$).
Here we choose to solve for $e^{-\sigma}=f_1(\Xp)$ and $e^{\, \sigma}=f_2(\Xm)$.
We can write down two arbitrary expansions for $e^{\pm \s}$,
\bea
\label{f1}
e^{-\s}=X_+ + g \left( \frac{\alpha}{2} X_+^{k+1} + \frac{\beta}{2}
X_+^{-k+1} \right) +g^2\left(B_1 X_+ +B_2 X_+^{2k+1} +B_3
X_+^{-2k+1}\right)+ \nonumber \\
+ g^3\left(B_4 X_+^{k+1} +B_5 X_+^{-k+1} +B_6 X_+^{3k+1}+B_7 X_+^{-3k+1} \right)
\ea
and
\bea
\label{f2}
e^{\, \s}=X_- + g \left( \frac{\alpha}{2} X_-^{-k+1} + \frac{\beta}{2}
X_-^{k+1} \right) +g^2\left(\tilde{B_1} X_- +\tilde{B_2} X_-^{-2k+1} +\tilde{B_3}
X_-^{2k+1}\right)+ \nonumber \\
+ g^3\left(\tilde{B_4} X_-^{-k+1} +\tilde{B_5} X_-^{k+1} +\tilde{B_6} X_-^{-3k+1}+\tilde{B_7} X_-^{3k+1}
\right).
\ea
The coefficients can be determined by plugging these ansatz into
(\ref{Xp}) and (\ref{Xm}), with $a(\s)$ given by (\ref{a}).
Solving perturbatively in $g$, one finds
\bea
B1&=&\frac{\alpha \beta}{2} + \frac{d_3}{8}, \; \; \; \;
B2=\frac{\alpha^2(1+k)}{4} + \frac{d_1}{8}, \; \; \; \;
B3=\frac{\beta^2(1-k)}{4} + \frac{d_2}{8},  \nonumber \\
B4&=& \alpha^2 \beta (\frac{k^2+5k+6}{16})
 + \alpha (\frac{2 d_3 + k d_3 }{16} )+ \beta (\frac{2 d_1 + k d_1}{16})-d_4 , \nonumber \\
B5&=&\alpha \beta^2 (\frac{k^2-5k+6}{16}) + \alpha (\frac{2 d_2 - k d_2 }{16} )+ \beta (\frac{2 d_3 - k d_3}{16})-d_5,    \nonumber \\
B6&=& \alpha^3(\frac{3k^2+5k+2}{16}) + \alpha(\frac{2d_1+3k d_1}{16})-d_6, \nonumber \\
B7&=& \beta^3(\frac{3k^2-5k+2}{16}) + \beta(\frac{2d_2-3k d_2}{16})-d_7,
\ea
and
\bea
\tilde{B1}&=&\frac{\alpha \beta}{2} + \frac{d_3}{8},  \; \; \;
\tilde{B2}= \frac{\alpha^2(1-k)}{4} + \frac{d_1}{8} \; \; \;
\tilde{B3}= \frac{\beta^2(1+k)}{4} + \frac{d_2}{8},  \nonumber \\
\tilde{B4}&=& \alpha^2 \beta (\frac{k^2-5k+6}{16})
 + \alpha (\frac{2 d_3 - k d_3 }{16} )+ \beta (\frac{2 d_1 - k d_1}{16})- d_4 ,  \nonumber \\
\tilde{B5}&=&\alpha \beta^2 (\frac{k^2+5k+6}{16}) + \alpha (\frac{2 d_2 + k d_2 }{16} )+ \beta (\frac{2 d_3 + k d_3}{16})-d_5 , \nonumber \\
\tilde{B6}&=&\alpha^3(\frac{3k^2-5k+2}{16}) + \alpha(\frac{2d_1-3k d_1}{16})-d_6, \nonumber \\
\tilde{B7}&=& \beta^3(\frac{3k^2+5k+2}{16}) + \beta(\frac{2d_2+3k d_2}{16})-d_7,
\ea
with
\bea
d_1 &=& \alpha^2(1 + 2k), \; \; \; d_2 = \frac{\beta^2}{\alpha^2}d_1,  \; \;
\; d_3 = 2\alpha \beta, \; \; \; d_4 = \alpha^2 \beta (\frac{-8c_2 -2k-3}{16}), \nonumber \\
d_5&=& \frac{\beta}{\alpha}d_4 , \; \; \; d_6 = \alpha^3(\frac{-8 c_1 -2k -1}{16}), \; \; \; d_7=\frac{\beta^3}{\alpha^3}d_6. \nonumber
\ea
Notice that $\tilde{B_i}=B_i(k \rightarrow -k)$, as could have
been seen directly from (\ref{f1}) and (\ref{f2}).
Finally, to find the Fermi surface, we simply plug into
\beq
X_+ X_- =a^2(e^{-\s},e^{\s})
\eq
the explicit expressions $e^{-\sigma}=f_1(\Xp)$ and $e^{\,
\sigma}=f_2(\Xm)$ given in (\ref{f1}) and (\ref{f2}).
We find that the Fermi surface, to cubic order in perturbation theory,
can be written as
\bea
X_+ X_- &=& 1-\frac{g}{2} \left[\alpha X_+^{k} + \alpha X_-^{-k} + \beta X_+^{-k} +\beta
X_-^{k}\right]+ \nonumber \\
&+& g^2 \left[-\frac{3k}{4} \left( \alpha^2 X_+^{2k}+\beta^2 X_-^{2k}\right) + \frac{k}{4} \left(\alpha^2 X_-^{-2k}
+\beta^2 X_+^{-2k} \right) \right]
+ \nonumber \\
&+& g^3 \left[(-\frac{5k^2}{4}-\frac{k}{4})(\alpha^3 X_+^{3k}+\beta^3
X_-^{3k})+(-\frac{3 k^2}{4}-\frac{k}{4})( \alpha^2 \beta X_+^{k}+\alpha
\beta^2 X_-^{k})\right]. \nonumber
\ea

It will turn out to be more convenient to rescale $k\rightarrow-k$
and to rewrite the surface in a slightly different way. After
using
$X_+ X_- = 1-\frac{g}{2} \left[\alpha X_+^{k} + \alpha X_-^{-k} + \beta X_+^{-k} +\beta
X_-^{k}\right]+ {\cal O}(g^2)$
to manipulate the quadratic and cubic terms, we get
\bea
\label{collFS}
X_+ X_- = 1-\frac{g}{2} \left[\alpha X_+^{-k} + \alpha X_-^k + \beta X_+^{k} +\beta X_-^{-k}\right]+
g^2 \frac{k}{2} \left[\alpha^2 X_+^{-2k}+\beta^2 X_-^{-2k}\right]
+ \nonumber \\
+ g^3 \left[(-\frac{3k^2}{4}+\frac{k}{4})(\alpha^3 X_+^{-3k}+\beta^3
X_-^{-3k})+(-\frac{k^2}{4}+\frac{k}{4})( \alpha^2 \beta X_+^{-k}+\alpha
\beta^2
X_-^{-k})\right].
\ea
Moreover, recall that the time dependence was incorporated into the
definition of $X_\pm$, $\Xp=e^{-t}(p+x)$ and $\Xm=e^{t}(p-x)$. We can make the time dependence explicit by
writing the surface in terms of $x$ and $p$ :
\bea
\label{TimeDepFS}
p^2-x^2 &=& 1-\frac{g}{2} \left[\alpha e^{kt} \Bigl((p+x)^{-k}+(p-x)^{k}\Bigr)+\beta e^{-kt}\Bigl((p+x)^{k}+(p-x)^{-k}\Bigr)
\right] + \nonumber \\
&+& g^2 \frac{k}{2} \Bigl[ \alpha^2 e^{2kt}(p+x)^{-2k} +\beta^2
e^{-2kt}(p-x)^{-2k} \Bigr]  + \nonumber \\
&+& g^3 (-\frac{3k^2}{4}+\frac{k}{4}) \Bigl[\alpha^3 e^{3kt}(p+x)^{-3k}  + \beta^3
e^{-3kt}(p-x)^{-3k} \Bigr] \nonumber \\
&+& g^3 (-\frac{k^2}{4}+\frac{k}{4}) \Bigl[ \alpha^2 \beta e^{kt}(p+x)^{-k} +\alpha
\beta^2 e^{-kt}(p-x)^{-k} \Bigr].
\ea

%%%%%%%%%%%%%%%%%%%%%%%%%%%%%%%%%%%%%%%%%%%%%%%%%%%%%%%%%%%%%%%%%%%%%%%%%%
\section{Another approach to building backgrounds}
%%%%%%%%%%%%%%%%%%%%%%%%%%%%%%%%%%%%%%%%%%%%%%%%%%%%%%%%%%%%%%%%%%%%%%%%%%

Another approach for constructing non-trivial backgrounds and their corresponding Fermi
surfaces was developed by AKK \cite{Alexandrov:2002fh,Kostov:2002tk}. It
is based on the idea of associating a new, perturbed Hamiltonian to each
deformed Fermi surface. They found equations for the
shape of the Fermi sea for a generic tachyon perturbation.
It will be visible that this approach is very different from that
of collective field theory.
For the special case of the Sine-Liouville background, we will
solve their equations, and find the explicit form of the Fermi
surface. We will compare it to the Fermi surface that we obtained in
the previous section, via collective field theory.
It is non-trivial to show that the two approaches are equivalent.

We start by outlining the construction used in \cite{Alexandrov:2002fh,Kostov:2002tk}.
Our calculations are then summarized in Sect. \ref{sub}. %%%%%%%%%%%%%%%%%%%%%%
Before introducing the concept of a deformed Fermi surface, recall
that the Fermi sea can be viewed as a collection of
classical trajectories having energies less than some fixed, Fermi
energy $E_F$. The ground state corresponds to the trajectory of
the most energetic fermion, with energy $E_F$.
For the standard inverted harmonic oscillator Hamiltonian
\beq
\label{H0}
H_0=-\frac{1}{2}(\hat{X}_+ \hat{X}_- + \hat{X}_- \hat{X}_+)
\eq
the profile of the Fermi sea is
\beq
\Xp \Xm=\mu,
\eq
where the Fermi energy was chosen to be $E_F=-\mu$.

Collective excitations are represented by deformations of
the Fermi surface, which can be obtained by replacing $\mu$ by a
general function of $X_\pm$,
\beq
\Xp \Xm= M(\Xp,\Xm).
\eq
While the ground state is stationary, the excited state
corresponding to such a deformed surface will generically be
time dependent.
If we want to think of this perturbation as a new fermion ground
state, we need to modify the one-fermion wave function to
incorporate the non-trivial features of the deformed Fermi
profile.
Before discussing the perturbed wave function, it is useful to
first discuss the case of the standard, inverted oscillator
potential.

%%%%%%%%%%%%%%%%%%%%%%%%%%%%%%%%%%%%%%%%%%%%%%%%%%%%%%%%%%%%%%%%%%%%%%%%%%%%%%%%

General solutions of the Schrodinger equation with Hamiltonian
(\ref{H0}), with a given energy $E$, take the form
$\psi^E_\pm(X_\pm,t)=e^{-iEt}\, \psi^E_\pm(X_\pm)$, with
\beq
\psi^E_\pm(X_\pm)=\frac{1}{2\pi}X_\pm^{\pm iE-1/2}.
\eq
These functions obey standard completeness and orthonormality
relations,
\bea
&& \langle \psi^E_\pm | \, \psi^{E'}_\pm \rangle
 = \int_0^\infty dX_\pm \overline{\psi^E_\pm(X_\pm)} \,
 \psi^{E'}_\pm(X_\pm) = \delta(E-E'), \nonumber \\
&&\int_{-\infty}^\infty  dE \; \overline{\psi^E_\pm(X_\pm)} \,
 \psi^{E}_\pm(X_\pm ') = \delta(X_\pm-X_\pm').
\ea
Moreover, the $\Xp$ and $\Xm$ representations can be
related to each other by a unitary operator $\hat{S}$,
\beq
[\hat{S} \, \psi_+](\Xm) = \int_0^\infty d\Xp K(\Xm,\Xp) \psi_+
(\Xp),
\eq
where the Kernel $K$ can be chosen to be
$$K(\Xm,\Xp)=\sqrt{\frac{2}{\pi}} \cos(\Xm\Xp).$$
Notice that since the operator $\hat{S}$ relates incoming and outgoing waves,
it can be interpreted as the fermionic scattering
matrix. To see this more explicitly, note that
\beq
[\hat{S}^\pm \, \psi^E_\pm](X_\pm)=\frac{1}{\pi} \int_0^\infty
dX_\pm \cos(\Xp\Xm) \, X_\pm^{\pm iE -1/2} = R(\pm E)
\psi^E_\pm(X_\mp),
\eq
with
\beq
R(E) = \sqrt{\frac{2}{\pi}} \cosh\left(\frac{\pi}{2}(\frac{i}{2}-E)\right) \, \Gamma(iE+1/2).
\eq
The factor $R(E)$, which is a pure phase, is the reflection
coefficient for scattering off the inverted oscillator potential.
Thus, scattering amplitudes between in and out states are given by
\beq
\langle \psi_- | \, \hat{S} \psi_+ \rangle = \langle \psi_- | \,K | \psi_+ \rangle
\eq
and in and out eigenfunctions satisfy the orthogonality relation
\beq
\langle \psi_-^E | \,K | \psi_+^{E'} \rangle = R(E) \delta(E-E').
\eq

So far we have restricted ourselves to the wave function of the
standard inverted oscillator Hamiltonian $H_0$.
Next, we would like to generalize these arguments to less trivial backgrounds.
If one wants to consider a perturbation such as $\Xp \Xm= M(\Xp,\Xm)$ as a new fermion
{\it ground} state, the wave function needs to be modified.
The perturbed wave functions $\Psi^E_\pm(X_\pm)$ can be related to the old, unperturbed
ones by a phase factor,
\beq
\label{newwavefn}
\Psi^E_\pm (X_\pm) = e^{\pm i \varphi_\pm (X_\pm,E)}
\psi^E_\pm(X_\pm).
\eq
The authors of \cite{Alexandrov:2002fh,Kostov:2002tk} parametrize
the phase in the following way:
\beq
\varphi_\pm (X_\pm;E)= \frac{1}{2} \phi(E) + V_\pm (X_\pm) + v_\pm
(X_\pm;E).
\eq
As will be shown later, the function $V_\pm (X_\pm)$ is what will be responsible for fixing
the perturbation unambiguously.
Notice that the phase $\phi(E)$ is precisely the analog of the reflection coefficient $R(E)$ discussed
above, for the standard oscillator case. Here it has been explicitly incorporated in the
expression for the new phase $\varphi$, and can be found by requiring
\beq
\hat{S} \Psi^E_+ = \Psi^E_- .
\eq
Thus, the orthonormality of in and out eigenfunctions takes the
form
\beq
\label{Portho}
\langle \Psi_-^{E_+} | \, K | \Psi_+^{E_-} \rangle =  \delta (E_{+} - E_{-}),
\eq
and fixes the shape of the perturbed wave function.

This integral can be evaluated, at the quasiclassical level, by the
saddle point approximation.
Plugging in the wave functions, (\ref{Portho}) takes the form
\bea
&& \frac{1}{\pi \sqrt{2\pi}} \int d\Xm \Xm^{iE_--1/2} e^{-i\varphi_-(\Xm)}
\int d\Xp \Xp^{iE_+-1/2} e^{-i\varphi_+(\Xp)} e^{i\Xp\Xm} \nonumber \\
&=& \frac{1}{\pi \sqrt{2\pi}} \int d\Xm \Xm^{iE_--1/2} e^{-i\varphi_-(\Xm)}
\int d\Xp \Xp^{-1/2} e^{\ln{\Xp^{iE_+}} - i\varphi_+(\Xp)+ i\Xp\Xm}.
\ea
Minimizing the phase
\beq
\frac{\partial}{\partial \Xp} \bigl( \ln{\Xp^{iE}} - i \varphi_+ (\Xp) + i \Xp \Xm
\bigr) = 0
\eq
yields
\beq
\Xp\Xm=-E + \Xp \, \p_+\, \varphi_+(\Xp),
\eq
where we used the delta function $\delta (E_{+} - E_{-})$ and set
$E \equiv E_{+}=E_{-}$.
Similarly, the saddle point approximation in the $\Xm$
integral gives
\beq
\Xp\Xm=-E + \Xm \, \p_{-}\, \varphi_-(\Xm).
\eq
Recalling that $E=-\mu$ and combining the two equations above, we
get
\beq
\label{DefFS}
\Xp\Xm= \mu + X_\pm \, \p_{\pm} \, \varphi_\pm(X_\pm),
\eq
which is precisely a deformed Fermi surface.
Furthermore, notice that for the trivial case $\varphi_\pm =0$ we
recover the unperturbed Fermi surface,
\beq
\Xp\Xm=-E=\mu.
\eq
Thus, we have seen how the profile of the Fermi sea arises by imposing bi-orthogonality on the
wavefunctions. Next, we would like to show how such surfaces are connected
to deformed Hamiltonians.

The functions $\Psi_\pm^E(X_\pm)$ are no longer eigenfunctions of the standard oscillator (\ref{H0}).
If we introduce the perturbed Hamiltonians $H_\pm$, associated with the $X_\pm$ representations, as solutions to the
equation
\beq
\label{Hdef}
H_\pm=H_0^\pm+X_\pm \partial_{\pm}\varphi_\pm(X_\pm,H_\pm),
\eq
one can show that $\Psi_\pm^E(X_\pm)$ are eigenfunctions of $H_\pm$ with eigenvalue $E$.
To see this, simply notice that
\beq
H_0^+ \Psi_+(\Xp) = -\Xp \partial_{\Xp} \varphi_+ \Psi_+(\Xp) + E
\Psi_+(\Xp),
\eq
and similarly for $ \Psi_-(\Xm)$.

%%%%%%%%%%%%%%%%%%%%%%%%%%%%%%%%%%%%%%%%%%%%%%%%%%%%%%%%%%%%%%%%%%%%%%%%%%

As one can see from the form of (\ref{DefFS}) and (\ref{Hdef}), in
order to specify a given background one must fully specify the phase $\varphi_\pm$.
In particular, it is enough to specify the form of the potentials $V_\pm$.
For tachyon perturbations
with momenta $p_n=n/R$, such potentials take the form
\beq
V_\pm(X_\pm)= R \sum_{k \geq 1} t_{\pm k} X_\pm^{k/R}.
\eq
Such a perturbation is exactly solvable, since it is generated by a
system of commuting flows associated to the coupling constants
$t_{\pm k}$.
The resulting integrable structure is that of a constrained Toda
lattice hierarchy.
Here for the sake of simplicity we have chosen not to review this method, which is described in detail
in \cite{Alexandrov:2002fh,Kostov:2002tk}.
Instead, we have motivated the form of the Fermi surface
by showing how it is connected to the perturbed Hamiltonian and
corresponding eigenfunctions.

Assuming that the phase $\varphi_\pm$ can be expanded as a Laurent series
\cite{Alexandrov:2002fh},
\beq
\label{varphi}
\varphi_\pm(X_\pm,E)= \frac{1}{2}\, \phi(E) + R\sum_{k\geq1}
t_{\pm k} \, X_\pm^{k/R} - R \sum_{k\geq1} \frac{1}{k} \, v_{\pm k} \, X_\pm^{-k/R},
\eq
and using $\Xp\Xm= \mu + X_\pm \, \p_{\pm} \, \varphi_\pm$, we
find
\beq
\Xp\Xm = \sum_{k \geq 1} k \, t_{\pm k} \, X_\pm^{k/R} + \mu +  \sum_{k \geq 1}  v_{\pm
k} \, X_\pm^{-k/R}.
\eq
The unknown coefficients $v_{\pm k}$ and the phase $\phi$ are not independent, but are functions of $t_{\pm k}$ and
$\mu$ (fixed by the bi-orthogonality condition); we will explicitly show how to find them for the Sine-Liouville case.

%%%%%%%%%%%%%%%%%%%%%%%%%%%%%%%%%%%%%%%%%%%%%%%%%
\subsection{Fermi surface construction \label{sub}}
%%%%%%%%%%%%%%%%%%%%%%%%%%%%%%%%%%%%%%%%%%%%%%%%%

The Sine-Liouville background is obtained by perturbing with the
lowest couplings, $t_{\pm 1}$.
The Fermi surface (\ref{DefFS}) can then be written in the two equivalent ways
\bea
\label{FSKostov1}
X_+ X_- &=& \mu + t_1 X_+^p + \sum_{k\geq 1} v_k X_+^{-pk}, \\
\label{FSKostov2}
X_+ X_- &=& \mu + t_{-1} X_-^p + \sum_{k\geq 1} v_{-k} X_-^{-pk},
\ea
where we defined $p=1/R$.
Since we are interested in the comparison with the collective field theory Fermi
surface, we need to solve for $v_{\pm k}$.
To do so, we will use the convenient ansatz of AKK, given by
\bea
\label{canonical}
X_+&=&A \, \omega(1+t_{-1} B \, \omega^{-p}), \nonumber \\
X_-&=&A \, \omega^{-1}(1+t_{1} B \, \omega^{p}),
\ea
with $A=e^{\frac{1}{2} \p_{\mu}\phi(E)}$ and $B=A^{p-2}$.
Plugging (\ref{canonical}) into (\ref{FSKostov1}) and collecting equal powers
of $\omega$, one obtains equations for $v_{\pm k}$ as well as for
the phase. Here we will stop at cubic order in perturbation
theory, neglecting ${\cal O}(t_{\pm 1}^4)$ terms.
Collecting the ${\cal O}(\omega^0)$ terms of (\ref{FSKostov1})
one finds the equation for the phase $\phi$,
\beq
\mu e^{-\p_{\mu}\phi}-(1-p)t_1t_{-1} e^{-(2-p)\p_{\mu}\phi}=1, \nonumber
\eq
which appears in \cite{Alexandrov:2002fh,Kostov:2002tk}.

Expanding $A=e^{\frac{1}{2} \p_{\mu}\phi(E)}$ perturbatively, we find
\beq
A=e^{\frac{1}{2} \p_{\mu}\phi(E)}=\sqrt{\mu}+t_1t_{-1}\frac{p-1}{2}\mu^{p-3/2}+{\cal
O}(t^4). \nonumber
\eq
Collecting the $\omega^{-p}$ powers of (\ref{FSKostov1})
gives
\beq
v_1= \mu^p \, t_{-1} + t_{1}\, t_{-1}^2 \, \mu^{2p-2}\, \frac{p\,(p-1)}{2}+{\cal
O}(t^4), \nonumber
\eq
while $v_2$ is obtained by looking at the $\omega^{-2p}$ terms,
\beq
v_2= p \, t_{-1}^2 \, \mu^{2p-1} + {\cal O}(t^4). \nonumber
\eq
Finally, the $\omega^{-3p}$ terms yield
\beq
v_3= t_{-1}^3 \, \mu^{3p-2}\, \frac{p\,(3p-1)}{2}. \nonumber
\eq

Similarly, the coefficients $v_{-k}$ are found by plugging the
ansatz (\ref{canonical}) into (\ref{FSKostov2}).
This yields
\bea
v_{-1}&=& \mu^p \, t_{1} + t_{1}^2\, t_{-1}\, \mu^{2p-2}\, \frac{p\,(p-1)}{2}, \nonumber \\
v_{-2}&=& p \, t_{1}^2 \, \mu^{2p-1},  \nonumber \\
v_{-3}&=& t_{1}^3 \, \mu^{3p-2} \, \frac{p\,(3p-1)}{2}, \nonumber
\ea
from the $\omega^{p}$,$\omega^{2p}$ and $\omega^{3p}$ terms
respectively.

Having found the coefficients $v_{\pm k}$, we can add the two Fermi surfaces (\ref{FSKostov1}) and
(\ref{FSKostov2}) to get the following:
\bea
\label{FSKostov3}
\Xp \Xm &=& \mu + \frac{1}{2} \left(t_1 \Xp^p + t_{-1} \Xm^{p} + \mu^p t_1 \Xm^{-p}
+ \mu^p t_{-1} \Xp^{-p}\right)+\nonumber \\
&+& \frac{p}{2} \left(t_1^2\, \mu^{2p-1}\, \Xm^{-2p} + t_{-1}^2 \, \mu^{2p-1}\, \Xp^{-2p}
\right)+  \frac{p\,(p-1)}{4} \, \mu^{2p-2} \left(t_1 \,t_{-1}^2  \Xp^{-p}+ t_1^2 \,t_{-1}\Xm^{-p} \right)
 +\nonumber \\
 &+&  \frac{p\,(3p-1)}{4}\, \mu^{3p-2}\, \left(t_1^3 \Xm^{-3p} +t_{-1}^3 \Xp^{-3p}
 \right)+ {\cal O}(t^4).
\ea

%%%%%%%%%%%%%%%%%%%%%%%%%%%%%%
\section{Comparison}
%%%%%%%%%%%%%%%%%%%%%%%%%%%%%%

To make the comparison with the collective field theory surface
(\ref{collFS}) more explicit, we can rewrite (\ref{FSKostov3}) in a more suggestive form,
\bea
\Xp \Xm &=& \mu + \frac{1}{2} \sqrt{\mu}^{\,p} \left[t_1 \left(\frac{\Xp}{\sqrt{\mu}}\right)^p +
t_{-1} \left(\frac{\Xm}{\sqrt{\mu}}\right)^p +  t_1 \left(\frac{\Xm}{\sqrt{\mu}}\right)^{-p}
+ t_{-1} \left(\frac{\Xp}{\sqrt{\mu}}\right)^{-p}\right]+\nonumber \\
&+& \frac{p}{2}\sqrt{\mu}^{\,2p-2} \left[t_1^2\, \left(\frac{\Xm}{\sqrt{\mu}}\right)^{-2p} + t_{-1}^2 \, \left(\frac{\Xp}{\sqrt{\mu}}\right)^{-2p}
\right]+ \nonumber \\
&+& \sqrt{\mu}^{\,3p-4}\, \frac{p\,(p-1)}{4}\left[t_1 \,t_{-1}^2  \left(\frac{\Xp}{\sqrt{\mu}}\right)^{-p}+
t_1^2 \,t_{-1}\left(\frac{\Xm}{\sqrt{\mu}}\right)^{-p} \right] \nonumber  \\
&+& \sqrt{\mu}^{\,3p-4} \frac{p\,(3p-1)}{4}\, \left[t_1^3 \left(\frac{\Xm}{\sqrt{\mu}}\right)^{-3p} +t_{-1}^3 \left(\frac{\Xp}{\sqrt{\mu}}\right)^{-3p}
  \right] + {\cal O}(t^4).
\ea
Defining new, rescaled coordinates
\beq
x_+ = \frac{\Xp}{\sqrt{\mu}}, \; \; \; x_- = \frac{\Xm}{\sqrt{\mu}}, \nonumber
\eq
the above Fermi surface becomes
\bea
x_+ x_- &=& 1 + \frac{1}{2} \sqrt{\mu}^{\,p-2} \left(t_1 x_+^p + t_{-1} x_-^{p} +  t_1 x_-^{-p} + t_{-1} x_+^{-p}\right)+
\frac{p}{2} \sqrt{\mu}^{\,2p-4} \left(t_1^2\, x_-^{-2p} + t_{-1}^2 \,  x_+^{-2p}
\right) + \nonumber \\
&+& \sqrt{\mu}^{\,3p-6} \left[\frac{p\,(p-1)}{4} \left(t_1 \,t_{-1}^2  x_+^{-p}+ t_1^2 \,t_{-1}x_-^{-p}
\right)+ \frac{p\,(3p-1)}{4} \left(t_1^3 x_-^{-3p} +t_{-1}^3 x_+^{-3p}
 \right)\right]. \nonumber
\ea
Finally, letting $t_{-1}=-\alpha, t_1=-\beta$, it takes the form
\bea
x_+ x_- &=& 1 - \frac{1}{2} \left(\sqrt{\mu}^{\,p-2}\right) \left(\alpha x_-^{p}+ \alpha x_+^{-p} + \beta x_+^p +\beta x_-^{-p} \right)+
\frac{p}{2}  \left(\sqrt{\mu}^{\,p-2}\right)^2 \left( \alpha^2 \,  x_+^{-2p} + \beta^2\, x_-^{-2p}\right) + \nonumber \\
&+&  \left(\sqrt{\mu}^{\,p-2}\right)^3 \left[\frac{p\,(-p+1)}{4} \left(\beta \,\alpha^2  x_+^{-p}+ \beta^2 \,\alpha x_-^{-p}
\right)+ \frac{p\,(-3p+1)}{4} \left(\alpha^3 x_+^{-3p} +\beta^3 x_-^{-3p}
\right)\right],\nonumber
\ea
which exactly matches the collective field theory surface
(\ref{collFS}) provided one identifies $g$ with $\sqrt{\mu}^{\,p-2}$ and $k$ with $p$.
Thus, we have seen that the collective field construction,
which starts from the standard inverted oscillator potential,
yields the same Fermi sea profile (to cubic order in perturbation theory) as the method of AKK, which on
the other hand made use of perturbed Hamiltonians.

\section{Conclusions}

A lot of recent research has been focusing on the construction of
time-dependent backgrounds in 2D string theory. Some of these non-trivial backgrounds
have been obtained by replacing the standard Liouville interaction
by a more general momentum or winding perturbation. In particular,
the Sine-Liouville background, which is a simple example of a momentum
perturbation, has received significant attention.
The Sine-Liouville model has been studied in two apparently different
ways;
one is based on the collective field theory description
of 2D string theory, while the other uses the tools of Toda
integrable systems.
While in the collective field approach
the Hamiltonian is that of the standard inverted harmonic
oscillator, the second method models the tachyon perturbation by introducing certain
deformed Hamiltonians.
In this work we have analyzed the two approaches to obtaining the Sine-Liouville background,
and we have constructed
in each case the explicit form of the Fermi surface.
Comparison of the two resulting surfaces demonstrates agreement
between the seemingly distinct methods, and sheds some light on the connection between them.
We would like to note that the collective field method allows
for the construction of non-trivial backgrounds
without the need to introduce
deformed Hamiltonians, at least for the types of perturbations considered here.
The issue of which backgrounds can be generated by the introduction of
appropriate non-trivial Hamiltonians is still an interesting one, which
we would like to further investigate.

\begin{acknowledgments}
I would like to thank Antal Jevicki for assistance throughout
this work, and Sumit Das for useful discussions.
Finally, I would like to thank G. L. Alberghi and S. Watson
for reading this manuscript.
This work was supported in part by the US Department of Energy under Contract DE-FG02-91ER40688, TASK A.
\end{acknowledgments}


\begin{thebibliography}{99}

%%%%%%%%%%%%%%%%%%% Reviews %%%%%%%%%%%%%%%%%%%%%%%%%%%%%%%%%%%%%%%%%%%%%%%%%%%%%%%%%%
\bibitem{Maldacena:1997re}
  J.~M.~Maldacena,
  ``The large N limit of superconformal field theories and supergravity,''
  Adv.\ Theor.\ Math.\ Phys.\  {\bf 2}, 231 (1998)
  [Int.\ J.\ Theor.\ Phys.\  {\bf 38}, 1113 (1999)] [arXiv:hep-th/9711200].

\bibitem{Klebanov:1991qa}
  I.~R.~Klebanov, ``String theory in two-dimensions,'' arXiv:hep-th/9108019.

\bibitem{Ginsparg:1993is}
  P.~H.~Ginsparg and G.~W.~Moore, ``Lectures on 2-D gravity and 2-D string theory,'' arXiv:hep-th/9304011.

\bibitem{Jevicki:1993qn}
  A.~Jevicki, ``Development in 2-d string theory,'' arXiv:hep-th/9309115.

\bibitem{Polchinski:1994mb}
  J.~Polchinski, ``What is string theory?,'' arXiv:hep-th/9411028.

\bibitem{Martinec:2004td}
  E.~J.~Martinec, ``Matrix models and 2D string theory,'' arXiv:hep-th/0410136.

%%%%%%%%%%%%%%%%%%%%%%%%%%%%%%%%%%%%%%%%%%%%%%%%%%%%%%%%%%%%%%%%%%%%%%%%%%%%%%%%%%%%%%%%
%\bibitem{Jevicki:1979mb}
%  A.~Jevicki and B.~Sakita,
%  ``The Quantum Collective Field Method And Its Application To The Planar Limit,''   Nucl.\ Phys.\ B {\bf 165}, 511 (1980).

\bibitem{McGreevy:2003kb}
  J.~McGreevy and H.~L.~Verlinde, ``Strings from tachyons: The c = 1 matrix reloaded,'' JHEP {\bf 0312}, 054 (2003)
  [arXiv:hep-th/0304224].

\bibitem{Klebanov:2003km}
  I.~R.~Klebanov, J.~Maldacena and N.~Seiberg, ``D-brane decay in two-dimensional string theory,''
  JHEP {\bf 0307}, 045 (2003) [arXiv:hep-th/0305159].

\bibitem{Sen:2003iv}
  A.~Sen,
  ``Open-closed duality: Lessons from matrix model,''
  Mod.\ Phys.\ Lett.\ A {\bf 19}, 841 (2004) [arXiv:hep-th/0308068].


\bibitem{Das:1990ka}
  S.~R.~Das and A.~Jevicki, ``String Field Theory And Physical Interpretation Of D = 1 Strings,''
  Mod.\ Phys.\ Lett.\ A {\bf 5}, 1639 (1990).

\bibitem{Polchinski:1991uq}
  J.~Polchinski, ``Classical limit of (1+1)-dimensional string theory,''
  Nucl.\ Phys.\ B {\bf 362}, 125 (1991).

%%%%%%%%%%%%%%%%%% Other backgrounds %%%%%%%%%%%%%%%%%%%%%%%%%%%%%%%%%%%%%%%%
\bibitem{Lee:1993fa}
  J.~Lee,
  ``Time dependent backgrounds in 2-d string theory and the S matrix generating functional,''
  Phys.\ Rev.\ D {\bf 49}, 2957 (1994)  [arXiv:hep-th/9310190].

\bibitem{Alexandrov:2002fh}
  S.~Y.~Alexandrov, V.~A.~Kazakov and I.~K.~Kostov,
  ``Time-dependent backgrounds of 2D string theory,'' Nucl.\ Phys.\ B {\bf 640}, 119 (2002)
  [arXiv:hep-th/0205079].

\bibitem{Kostov:2002tk}
  I.~K.~Kostov,
  ``Integrable flows in c = 1 string theory,'' J.\ Phys.\ A {\bf 36}, 3153 (2003)
  [Annales Henri Poincare {\bf 4}, S825 (2003)]  [arXiv:hep-th/0208034].

%\bibitem{Hikida:2004mp}
%  Y.~Hikida and T.~Takayanagi,
%  ``On solvable time-dependent model and rolling closed string tachyon,'' Phys.\ Rev.\ D {\bf 70}, 126013 (2004)
%  [arXiv:hep-th/0408124].

%\bibitem{Takayanagi:2004yr}
%  T.~Takayanagi, ``Matrix model and time-like linear dilaton matter,''
%  JHEP {\bf 0412}, 071 (2004) [arXiv:hep-th/0411019].

\bibitem{Karczmarek:2003pv}
  J.~L.~Karczmarek and A.~Strominger, ``Matrix cosmology,''
  JHEP {\bf 0404}, 055 (2004) [arXiv:hep-th/0309138].

\bibitem{Karczmarek:2004ph}
  J.~L.~Karczmarek and A.~Strominger, ``Closed string tachyon condensation at c = 1,''
  JHEP {\bf 0405}, 062 (2004) [arXiv:hep-th/0403169].

\bibitem{Das:2004hw}
  S.~R.~Das, J.~L.~Davis, F.~Larsen and P.~Mukhopadhyay,
  ``Particle production in matrix cosmology,''
  Phys.\ Rev.\ D {\bf 70}, 044017 (2004)
  [arXiv:hep-th/0403275].

\bibitem{Karczmarek:2004yc}
  J.~L.~Karczmarek, A.~Maloney and A.~Strominger, ``Hartle-Hawking vacuum for c = 1 tachyon condensation,''
  JHEP {\bf 0412}, 027 (2004) [arXiv:hep-th/0405092].

\bibitem{Ernebjerg:2004ut}
  M.~Ernebjerg, J.~L.~Karczmarek and J.~M.~Lapan, ``Collective field description of matrix cosmologies,''
  JHEP {\bf 0409}, 065 (2004) [arXiv:hep-th/0405187].

\bibitem{Mukhopadhyay:2004ff}
  P.~Mukhopadhyay,
  ``On the problem of particle production in c = 1 matrix model,''
  JHEP {\bf 0408}, 032 (2004)
  [arXiv:hep-th/0406029].

\bibitem{Das:2004aq}
  S.~R.~Das and J.~L.~Karczmarek, ``Spacelike boundaries from the c = 1 matrix model,''
  Phys.\ Rev.\ D {\bf 71}, 086006 (2005) [arXiv:hep-th/0412093].

\bibitem{Das:2005jp}
  S.~R.~Das, ``Non-trivial 2d space-times from matrices,''
  arXiv:hep-th/0503002.

\end{thebibliography}
\end{document}